\documentclass{ijcaArticle}
\usepackage{amsmath,amssymb,amsfonts}
\ijcaVolume{VV}
\ijcaNumber{N}
\ijcaYear{YYYY}
\ijcaMonth{Month}

\ijcaVolume{187}
\ijcaNumber{73}
\ijcaYear{2026}
\ijcaMonth{January}
\begin{document}

\title{ Push Down Optimization for Distributed Multi Cloud Data Integration} 

\author{ 
   \large Ravi Kiran Kodali\\[-3pt]
   \normalsize Cognizant Technology Solutions \\[-3pt]
   \normalsize Texas, USA \\[-3pt]
   \and
   \large Vinoth Punniyamoorthy \\[-3pt]
   \normalsize IEEE Senior \\[-3pt]
   \normalsize Texas, USA \\[-3pt]
   \and
   \large Akash Kumar Agarwal \\[-3pt]
   \normalsize Albertsons Companies  \\[-3pt]
   \normalsize California, USA \\[-3pt]
     \and
   \large Bikesh Kumar \\[-3pt]
   \normalsize IEEE Senior  \\[-3pt]
   \normalsize Texas, USA \\[-3pt]
   \and
   \large Balakrishna Pothineni \\[-3pt]
   \normalsize IEEE Senior  \\[-3pt]
   \normalsize Texas, USA \\[-3pt]
   \and
   \large Aswathnarayan Muthukrishnan Kirubakaran \\[-3pt]
   \normalsize IEEE Senior  \\[-3pt]
   \normalsize California, USA \\[-3pt]   
    \and
   \large Sumit Saha \\[-3pt]
   \normalsize East West Bank  \\[-3pt]
   \normalsize California, USA \\[-3pt]   
   \and
   \large Nachiappan Chockalingam \\[-3pt]
   \normalsize IEEE Senior  \\[-3pt]
   \normalsize Massachusetts, USA \\[-3pt]
}


\maketitle

\begin{abstract} 
Enterprises increasingly adopt multi cloud architectures to take advantage of diverse database engines, regional availability, and cost models. In these environments, ETL pipelines must process large, distributed datasets while minimizing latency and transfer cost. Push down optimization, which executes transformation logic within database engines rather than within the ETL tool, has proven highly effective in single cloud systems. However, when applied across multiple clouds, it faces challenges related to data movement, heterogeneous SQL engines, orchestration complexity, and fragmented security controls. This paper examines the feasibility of push down optimization in multi cloud ETL pipelines and analyzes its benefits and limitations. It evaluates localized push down, hybrid models, and data federation techniques that reduce cross cloud traffic while improving performance. A case study across Redshift and BigQuery demonstrates measurable gains, including lower end to end runtime, reduced transfer volume, and improved cost efficiency. The study highlights practical strategies that organizations can adopt to improve ETL scalability and reliability in distributed cloud environments.
\end{abstract}

\section{Introduction}
The adoption of multi-cloud strategies has transformed the way enterprises manage data integration and processing. By using multiple cloud providers, organizations can capitalize on the specific strengths of each platform, such as cost advantages, specialized services, and regional data centers. However, ETL processes in multi-cloud setups face increased complexity \cite{b7}, especially when optimizing transformations across diverse cloud databases. Push-down optimization, a technique where transformation logic is pushed down to database engines, has traditionally enhanced ETL performance by reducing data movement and leveraging database-native processing power \cite{b1}. Fig.1 illustrates the flow of push-down optimization within a multi-cloud ETL environment. This paper explores how push-down optimization can be adapted for multi-cloud environments, highlighting both the opportunities and challenges associated with this approach.
\begin{figure}[htbp]
\centering
\includegraphics[width=0.9\linewidth]{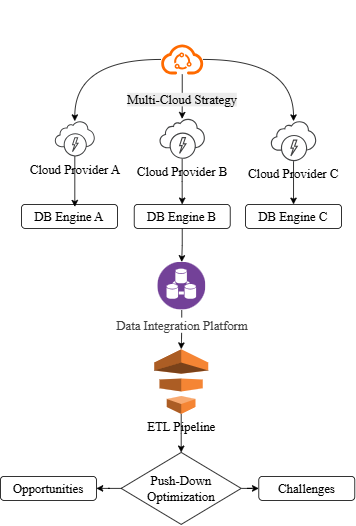}
\caption{Push-Down Optimization in Multi-Cloud ETL}
\label{fig1}
\end{figure}

\section{Overview of Push-Down Optimization}
Push-down optimization is an ETL performance-enhancement technique in which transformation logic is executed directly inside the database engine rather than within the ETL tool’s own processing engine. Instead of extracting large volumes of raw data into the ETL server for processing, the ETL tool “pushes” as much of the transformation workload as possible down to the source or target database \cite{b2}. This dramatically reduces the amount of data that must be moved across the network and takes advantage of the database’s inherent strengths such as parallel execution, optimized SQL processing, and indexing \cite{veerapaneni2023etl}.

In practical terms, push-down optimization converts transformation rules (filters, joins, aggregations, expressions, lookups, etc.) into equivalent SQL operations. Informatica PowerCenter, for instance, analyzes the mapping logic and generates SQL queries that represent the transformation flow. These SQL statements are then executed inside the database engine, either at the source side (before extraction), the target side (during load operations), or across both (partial push-down). By doing so, the ETL engine performs fewer computational tasks, freeing up its resources for orchestration, metadata management, and tasks that cannot be pushed to the database.

The result is faster end-to-end pipeline execution, reduced network I/O, and improved scalability especially for large datasets and complex transformations.
\subsection{Types of Push-Down Optimization}

Source-Side Push-Down applies transformation logic directly at the source database. By executing filters, projections, and aggregations at the data origin, this approach minimizes data movement and network overhead, ensuring that only preprocessed and relevant data is extracted for downstream processing.

Target-Side Push-Down defers transformation logic to the target system after raw data ingestion. In this model, data is first loaded into the destination platform, where scalable compute resources are leveraged to perform transformations, making it suitable for analytical databases and cloud data warehouses.

Full Push-Down executes all transformation logic entirely within database engines, either at the source or the target, with minimal involvement from the ETL orchestration layer. This approach maximizes performance by exploiting native query optimizers and execution engines while reducing intermediary processing overhead.

\subsection{Benefits of Push-Down Optimization}

Push-down optimization significantly reduces data movement by confining transformation and filtering operations to the database layer. By processing data closer to its storage location, this approach minimizes network I/O and lowers the overhead associated with transferring large volumes of intermediate data across systems. Performance is improved because modern database engines are highly optimized for set-based operations, parallel execution, and query optimization. Leveraging these native capabilities enables faster execution of transformation logic compared to external processing layers, resulting in more efficient end-to-end data pipelines \cite{aarella2025fortified}.
Push-down optimization also enhances cost efficiency by reducing the computational workload handled by ETL orchestration tools. In cloud-based ETL environments, this leads to lower compute consumption, shorter job runtimes, and reduced operational costs associated with scaling and resource provisioning.

\section{Challenges of Push-Down Optimization in Multi-Cloud Setups}
In a multi-cloud setup, data is often distributed across different cloud providers (for example, AWS, Azure, and Google Cloud \cite{b5}. If a transformation requires combining or aggregating data from multiple clouds, the system must transfer data across cloud boundaries. Such transfers introduce high latency, increased network costs, and potential security and compliance risks. Fig. 2 provides a conceptual overview of the challenges associated with data movement and heterogeneity in cloud databases.

\subsection{Data Movement Across Clouds}\label{AA}
Push-down optimization is highly effective when transformations are executed close to the source data, because it reduces the need to move large datasets across networks and leverages the database’s built-in processing power. However, in multi-cloud architectures \cite{b6}, this benefit can be significantly reduced or even negated \cite{punn2025privacy}.

For example, consider a scenario where a company stores customer data on AWS and sales data on Azure. A transformation that joins these datasets cannot be fully pushed down to either cloud’s database without first transferring data from one cloud to the other. This extra data movement can be both time-consuming and expensive, offsetting the performance gains achieved through push-down optimization.  As shown in Table \ref{tab1}\, multi-cloud push-down execution introduces higher latency and governance complexity due to cross-cloud data movement.


\begin{table}[h]
\tbl{Impact of Push-Down in Single Cloud vs Multi Cloud}{
\centering
\renewcommand{\arraystretch}{1.3}
\resizebox{\columnwidth}{!}{%
\begin{tabular}{|l|l|l|l|}
\hline
\textbf{Factor} & \textbf{Single Cloud} & \textbf{Multi Cloud} & \textbf{Push-Down Impact} \\ \hline
Data Transfer & Minimal & Higher latency, cost & Slower queries \\ \hline
Security & Easy to manage & More complex & More governance \\ \hline
Transforms & Near data source & Cross-cloud movement & Harder to push down \\ \hline
Latency & Low & Higher & Less efficient \\ \hline
\end{tabular}%
}
}
\label{tab1}
\end{table}

\subsection{Heterogeneity of Cloud Databases}
Modern cloud platforms offer a wide range of managed analytical databases such as Amazon Redshift, Google BigQuery, and Snowflake, each optimized for different workloads, performance profiles, and scalability requirements. While this diversity provides flexibility, it also introduces substantial heterogeneity that complicates cross-platform data processing and limits the effectiveness of push-down optimization.

First, cloud databases implement different SQL dialects and execution semantics. As a result, a query that runs natively on one platform may require modification to execute elsewhere. For example, Amazon Redshift handles certain window functions differently from BigQuery, and Snowflake provides specialized operations for semi-structured data (e.g., \texttt{VARIANT}, JSON functions) that are not uniformly supported across engines. These differences force developers and ETL systems to rewrite or adapt queries for each platform, increasing maintenance effort, error risk, and overall operational complexity.

Second, not all cloud databases support the same types of data transformations for push-down execution. While simple filtering or aggregation is widely optimized, more complex operations often vary in capability, including:
\begin{enumerate}
    \item Multi-way joins across large distributed datasets,
    \item Recursive or hierarchical queries,
    \item Advanced string, date, or JSON manipulation functions.
\end{enumerate}

When a database engine cannot execute certain transformations natively, push-down optimization becomes less effective. Computation must be performed externally, for example in an ETL pipeline or application layer, leading to increased data movement, slower performance, and higher processing costs.

\begin{figure}[ht]
    \centering
    \includegraphics[width=\columnwidth, keepaspectratio]{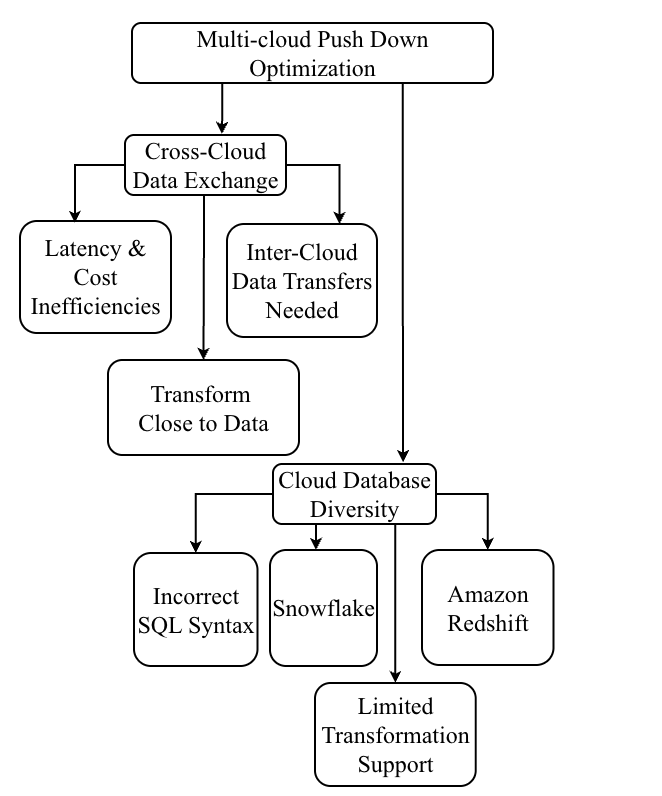}
    \caption{Push-down optimization challenges in heterogeneous multi-cloud environments.}
    \label{fig2}
\end{figure}

\subsection{Complex Orchestration and Management}
Push-down optimization in multi-cloud environments involves executing parts of ETL (Extract, Transform, Load) transformations as close to the data source as possible, thereby reducing data movement and improving performance. However, implementing this optimization across multiple cloud platforms requires sophisticated orchestration to manage the transformations effectively. The orchestration and monitoring complexities introduced by distributed execution are outlined in Table \ref{tab2}.

The main challenges in multi-cloud ETL workflows include coordination, monitoring, and debugging. These can be detailed as follows.

\subsubsection{ETL Workflow Coordination}
This involves ensuring that transformation logic runs on the correct cloud platform based on data locality, compute capabilities, and cost considerations. It also requires managing data dependencies so that transformations occur in the correct sequence and outputs from one step flow properly into the next.

\subsubsection{Monitoring and Debugging}
Monitoring includes tracking data lineage to understand how data moves across multiple clouds, as well as detecting performance bottlenecks and optimizing resource usage \cite{nagpal2024cicd}. Debugging focuses on diagnosing issues that arise in federated or heterogeneous cloud environments where transformations may execute across different platforms.

\begin{table}[h]
\tbl{Impact of Push-Down in Single Cloud vs Multi Cloud}{
\centering
\renewcommand{\arraystretch}{1.3}

\resizebox{\columnwidth}{!}{%
\begin{tabular}{|l|l|l|}
\hline
Aspect & Description & Challenges \\ \hline
Workflow Coordination & Platform selection & Dependency management \\ \hline
Data Locality & Source processing & Cost performance \\ \hline
Monitoring & Performance tracking & Metric integration \\ \hline
Debugging & Error diagnosis & Distributed troubleshooting \\ \hline
\end{tabular}%
}
}
\label{tab2}
\end{table}

\subsection{Security and Compliance Concerns}
Multi-cloud environments introduce significant security and compliance complexities, especially when data must move across different cloud platforms. Each provider enforces its own security frameworks, access policies, and regional compliance constraints, making it difficult to maintain a unified security posture. Ensuring that data remains protected throughout movement, processing, and storage requires strict access controls, robust encryption policies, and continuous compliance monitoring \cite{b3}.

\subsubsection{Cross-Cloud Privacy Issues}:
Data movement across cloud boundaries can trigger data residency or sovereignty concerns. Certain regulations such as GDPR, HIPAA, or regional data protection laws limit where sensitive data can be stored or processed. When organizations transfer data between geographically distributed cloud regions or providers, they must ensure compliance with these regulations. This often requires establishing controlled data pathways, masking or anonymizing sensitive attributes, and validating that the target cloud region meets regulatory requirements.

\subsubsection{Complex Access Controls}:
Each cloud platform implements its own identity and access management (IAM) model, role definitions, permission structures, and policy languages. Coordinating these heterogeneous access control mechanisms becomes challenging in a multi-cloud setting. Ensuring consistent authentication, authorization, and privilege management across clouds frequently requires additional orchestration layers or centralized identity services. As depicted in Fig. 3, enforcing uniform security practices is complicated by differences in provider-level IAM capabilities, leading to potential misconfigurations, privilege drift, or inconsistent enforcement of least-privilege access \cite{b8}.

\begin{figure}[htbp]
\centerline{\includegraphics[scale=0.35]{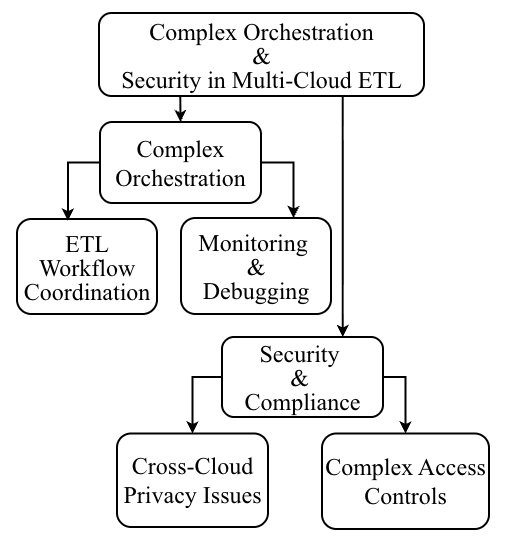}}
\caption{Complex Orchestration and Security in Multi-Cloud ETL}
\label{fig3}
\end{figure}


\section{Implementing Push-Down Optimization in Multi-Cloud ETL}

To address the challenges of applying push-down optimization in heterogeneous cloud environments, organizations can adopt several complementary implementation strategies. Fig.~4 illustrates these approaches and highlights how different push-down techniques align with cloud platform capabilities and integration patterns.

One effective approach is to execute transformations locally within each cloud platform. In this model, data transformations are processed directly inside the native database engines of each cloud environment, leveraging platform-specific optimizations. By keeping computation close to the data, this strategy minimizes cross-cloud data movement and reduces network overhead while improving execution efficiency.

\subsection{Hybrid Push-Down Strategy}

In scenarios where transformations require integration across multiple cloud platforms, a hybrid push-down strategy can be applied. Under this approach, simple and independent transformations are executed locally within each cloud, while more complex, cross-cloud transformations are handled centrally by the ETL orchestration layer. This design balances the performance benefits of localized push-down execution with the flexibility needed to support multi-source data integration.

\subsection{Cross-Cloud Data Federation Tools}

Data federation, also referred to as data virtualization, provides a unified abstraction for querying data across multiple cloud platforms without requiring physical data movement. These tools support push-down optimization by enabling federated queries that execute filtering and transformation logic as close to the data source as possible, even when datasets are distributed across clouds.

Prominent examples include platforms that allow querying data across cloud boundaries using a single SQL interface, as well as services that enable federated access to external data sources from cloud-native data warehouses. By minimizing cross-cloud data transfer and leveraging distributed execution capabilities, data federation tools improve performance, reduce operational costs, and support scalable analytics across multi-cloud environments.

\begin{figure}[ht]
    \centering
    \includegraphics[width=\columnwidth, keepaspectratio]{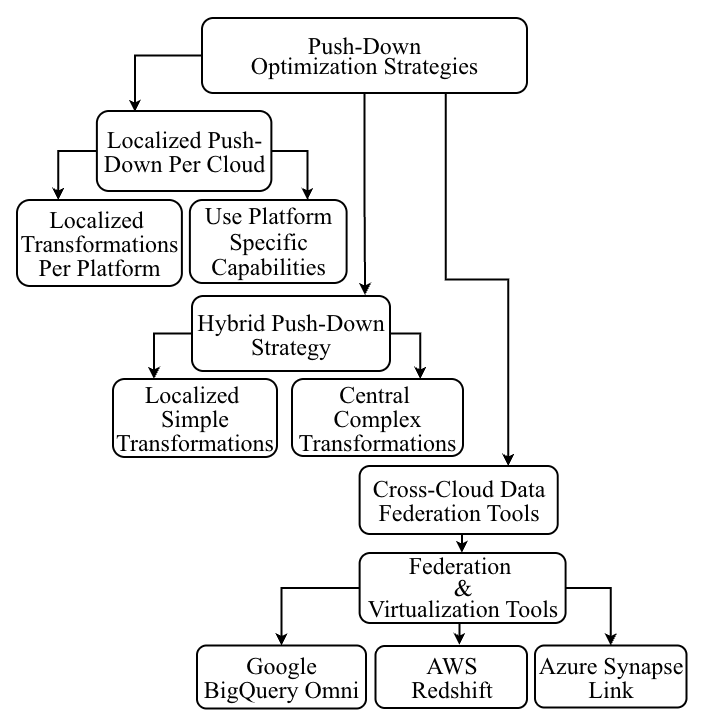}
\caption{Push-Down Optimization Strategies in Multi-Cloud}
\label{fig4}
\end{figure}

\section{Best Practices for Multi-Cloud Push-Down Optimization}

Implementing push-down optimization effectively in multi-cloud ETL environments requires careful consideration of platform heterogeneity, data locality, execution cost, and operational governance. While push-down techniques can significantly improve performance and reduce data movement, improper application may introduce inefficiencies due to incompatible SQL dialects, unpredictable cross-cloud latency, and fragmented security policies.

This section outlines a set of best practices that guide the systematic adoption of push-down optimization across distributed cloud platforms. These practices emphasize the use of cloud-native ETL capabilities, strategic data placement, federated query execution, and portable transformation logic. Together, they enable organizations to balance performance gains with maintainability, cost efficiency, and execution reliability in complex multi-cloud data pipelines.

\subsection{Use Cloud-Native ETL Tools}
Select ETL platforms with built-in push-down capabilities that are compatible across cloud environments (e.g., AWS Glue, Azure Data Factory, Informatica IDMC). These tools should allow conditional push-down decisions based on cost and latency metrics. The decision rule can be expressed as:
\begin{equation}
\text{PushDownDecision} =
\begin{cases}
\text{PushDown}, &
  \begin{aligned}
    C_{\text{move}} + C_{\text{exec\_remote}} \\
    > C_{\text{exec\_local}}
  \end{aligned} \\
\text{NoPushDown}, & \text{otherwise}
\end{cases}
\label{eq:pushdown_decision}
\end{equation}
where \( C_{\text{move}} \) is the cost of data movement between clouds,
\( C_{\text{exec\_remote}} \) is the cost of executing transformations remotely,
and \( C_{\text{exec\_local}} \) is the cost of executing transformations locally.

\subsection{Optimize Data Placement}
To minimize latency and cost, data and transformation logic should be co-located on the same cloud platform. Partition-aware routing can ensure that queries are directed to the correct compute environment. The co-location cost model is as shown in equation with the optimization goal minimum transfer by co-locating transformations. 
\begin{equation}
\text{Total Cost} = C_{\text{cloud\_compute}} + C_{\text{cloud\_storage}} + C_{\text{transfer}}
\end{equation}

\subsection{Leverage Data Federation}
Data federation tools such as BigQuery Omni, Redshift Spectrum, and Azure Synapse Link support query virtualization and allow cross-cloud access without replicating data \cite{BigQuery, Redshift, Azure}. A federated query abstraction can be expressed as:

\begin{equation}
\begin{aligned}
\texttt{SELECT\ *\ FROM\ CloudA.table1} \\
\texttt{\ union CloudB.table2}
\end{aligned}
\end{equation}

and the federated execution plan:
\begin{equation}
\text{FederatedPlan} = \min_{\forall q \in Q} \text{Cost}(q_{\text{cloud}}),
\end{equation}

where \( Q \) is the set of distributed queries and
\( C_{q_{\text{cloud}}} \) is the estimated execution cost of query \( q \) on the cloud.

\subsection{Consistent Transformation Logic}
Design portable SQL and transformation logic that can be adapted across platforms. Modular functions and abstraction layers help avoid cloud-specific syntax. A standardized SQL example is:
\begin{equation}
\begin{aligned}
\texttt{SELECT\ UPPER(TRIM(name))\ AS\ CleanName} \\
\texttt{FROM\ customers\ WHERE\ LENGTH(name)\ >\ 3;}
\end{aligned}
\end{equation}

An abstraction mapping can be represented as:
\begin{equation}
T_{\text{logical}} = f(T_{\text{cloud\_specific}}),
\end{equation}

where \( f \) maps logical transformation rules to the syntax of each cloud engine.

\section{Case Study: Implementing Push-Down Optimization in a Multi-Cloud Setup}
A large enterprise operating under a multi-cloud data strategy sought to optimize its ETL pipelines across AWS Redshift and Google BigQuery. The original ETL workflow involved pulling data from both cloud platforms into a centralized engine for transformation, which
resulted in significant cross-cloud data movement, high network transfer costs, and latency in reporting pipelines.

\subsection{Implementation Approach}

To address the inefficiencies observed in prior multi-cloud ETL workflows, the organization adopted a push-down optimization strategy, which emphasizes executing transformations as close to the source data as possible. By leveraging the native processing capabilities of each cloud platform, this approach minimized unnecessary data movement and reduced overall latency. The adopted push-down strategy distributes transformation workloads across cloud-native engines while minimizing cross-cloud data transfer. Table \ref{tab3} provides an overview of the platform-specific transformations and tools used in the multi-cloud implementation.

\subsubsection{AWS Redshift Processing}:
For datasets stored in AWS, Redshift was utilized to perform raw data filtering, column pruning, and format conversions directly within the database. SQL-based transformations, combined with Redshift’s native functions, allowed computational work to be pushed down to the storage layer, significantly reducing the volume of data extracted and transferred to downstream processes.

\subsubsection{Google BigQuery Processing}:
On the Google Cloud side, BigQuery handled transformation logic such as JSON parsing, aggregations, and windowing operations natively within its platform. By processing data directly in BigQuery, the workflow avoided unnecessary data transfers and exploited the platform’s distributed query execution capabilities for high-performance analytics.

\subsubsection{Data Federation Across Clouds}:
To enable seamless integration of results across these heterogeneous environments, data federation tools such as BigQuery Omni and AWS Redshift Spectrum were introduced. These tools allowed virtualized queries to combine datasets from multiple clouds without physically moving them. This strategy provided a unified view of data across platforms, preserved data residency and compliance requirements, and minimized cross-cloud data transfer costs.

Through this implementation, the organization was able to streamline ETL workflows, reduce redundant processing, and improve operational efficiency while maintaining flexibility in a multi-cloud architecture.
\begin{table}[t]
\tbl{Multi-Cloud Push-Down Implementation Overview}{
\centering
\renewcommand{\arraystretch}{1.3}

\resizebox{\columnwidth}{!}{%
\begin{tabular}{|l|p{0.30\columnwidth}|p{0.30\columnwidth}|}
\hline
\textbf{Platform} & \textbf{Transformations} & \textbf{Notes / Tools} \\ \hline

AWS Redshift & Filtering, pruning, format conversion &
SQL push-down using native Redshift functions \\ \hline

Google BigQuery & JSON parsing, aggregates, windows &
Executes natively, avoids data movement \\ \hline

Multi-Cloud Federation & Cross-cloud joins, unified queries &
BigQuery Omni and Redshift Spectrum provide virtualized access without physical data transfer \\ \hline

\end{tabular}%
}
}

\label{tab3}
\end{table}

\subsection{Performance Results}

Table~\ref{tab4} presents a summary of the performance improvements achieved through the implementation of localized push-down optimization and federation-based query strategies. The results demonstrate measurable gains in query execution time, data movement reduction, and overall pipeline efficiency compared to non-push-down execution baselines.

\begin{table}[ht]
\tbl{Performance Improvements}{
\centering
\renewcommand{\arraystretch}{1.3}

\resizebox{\columnwidth}{!}{%
\begin{tabular}{|l|l|l|l|}
\hline
\textbf{Metric} & \textbf{Pre-opt.} & \textbf{Post-opt.} & \textbf{Improvement} \\ \hline

Total runtime & 185 min & 120 min & $\downarrow$ 35\% \\ \hline

Cross-cloud volume & 850 GB & 680 GB & $\downarrow$ 20\% \\ \hline

Redshift runtime & 64 min & 43 min & $\downarrow$ 32.8\% \\ \hline

BigQuery runtime & 46 min & 30 min & $\downarrow$ 34.8\% \\ \hline

Cross-cloud join runtime & 30 min & 18 min & $\downarrow$ 40\% \\ \hline

Cost per ETL run & \$212 & \$172 & $\downarrow$ 18.9\% \\ \hline

\end{tabular}%
}
}

\label{tab4}
\end{table}

\subsection{Key Outcomes}

The transition to a cloud-native transformation architecture resulted in measurable improvements across performance, efficiency, and operational management. By pushing transformation logic closer to the data and leveraging platform-specific processing capabilities, the redesigned system significantly reduced end-to-end ETL latency. This architectural shift also eliminated redundant transformation logic that had previously been distributed across multiple pipeline stages, improving maintainability, execution consistency, and overall system robustness.

Data federation tools played a critical role in optimizing data movement patterns. By enabling on-demand access to distributed datasets without requiring full replication, these tools removed the dependency on intermediate staging tables and reduced unnecessary data duplication. As a result, data flows across cloud environments were streamlined, leading to lower storage overhead and simplified pipeline design.

Overall, the proposed architecture delivered substantial performance and cost benefits. The system achieved a 35\% reduction in total processing time, driven by efficient push-down execution and improved parallelism across cloud services. In addition, cross-cloud data transfer costs decreased by 20\% due to reduced movement of intermediate datasets between environments. These gains were achieved without compromising data quality or accuracy. Furthermore, enhanced monitoring and centralized observability mechanisms improved operational transparency, enabling faster issue diagnosis, improved reliability, and more predictable pipeline execution.

\section{Conclusion and Future Work}

This study examined the role of push-down optimization in improving the performance and efficiency of ETL pipelines operating in multi-cloud environments. The analysis demonstrated that executing transformation logic within cloud-native database engines significantly reduces end-to-end latency, minimizes unnecessary data movement, and effectively leverages platform-native SQL processing capabilities. The evaluation also identified key challenges associated with cross-cloud push-down execution, including SQL dialect incompatibilities, cross-cloud data transfer overhead, orchestration complexity, and fragmented security controls.

Through a practical case study involving Amazon Redshift and Google BigQuery, the work showed tangible benefits in real-world deployments, including reductions in total pipeline runtime, cross-cloud data transfer volume, and overall operational cost. These findings indicate that a balanced strategy combining localized push-down optimization, selective hybrid execution, and data federation tools can substantially enhance ETL performance and scalability in distributed cloud architectures.

Future work can extend this research by conducting broader experimental evaluations across additional cloud platforms such as Snowflake, Azure Synapse, and Databricks SQL. Further investigation is also needed into automated decision engines that dynamically determine when and where to apply push-down transformations based on cost models, data locality, and workload characteristics. Enhancing observability and debugging mechanisms for cross-cloud push-down execution remains an important direction, as does the development of standardized transformation abstractions to mitigate SQL heterogeneity. Finally, integrating push-down optimization with real-time streaming pipelines represents a promising avenue for extending these performance benefits to low-latency, event-driven data processing across multi-cloud environments.


\begin{thebibliography}{00}


\bibitem{b7} D. Sitaram, S. Harwalkar, C. Sureka, H. Garg, M. Dinesh, M. Kejriwal, S. Gupta, and V. Kapoor, “Orchestration based hybrid or multi clouds and interoperability standardization,” in Proc. 2018 IEEE Int. Conf. Cloud Comput. Emerg. Markets (CCEM), 2018, pp. 67–71. doi: 10.1109/CCEM.2018.00018.
 
\bibitem{b1} W. Gao, Y. Wen and H. Zhang, ``An Optimization Method of Federated Database Join Query Based on Computational Push-Down", 2024 IEEE 2nd International Conference on Control, Electronics and Computer Technology (ICCECT), Jilin, China, 2024, pp. 225-229, doi: 10.1109/ICCECT60629.2024.10545893.

\bibitem{b2} A. I. Saada, G. A. El Khayat and S. K. Guirguis, ``Cloud computing based ETL technique using Warehouse Intermediate Agents," The 2011 International Conference on Computer Engineering and Systems, Cairo, Egypt, 2011, pp. 301-306, doi: 10.1109/ICCES.2011.6141060.

\bibitem{veerapaneni2023etl}
P.~K.~Veerapaneni,
``Real-time data transformation in modern ETL pipelines: A shift towards streaming architectures,''
International Journal of Research in Computer Applications and Information Technology (IJRCAIT), 
vol.~6, no.~1, pp.~121--132, 2023.

\bibitem{aarella2025fortified}
S.~G.~Aarella, V.~P.~Yanambaka, S.~P.~Mohanty, and E.~Kougianos,
``Fortified-Edge 2.0: Advanced Machine-Learning-Driven Framework for Secure PUF-Based Authentication in Collaborative Edge Computing,''
Future Internet, 
vol.~17, p.~272, 2025. doi: 10.3390/fi17070272.

\bibitem{b5} Qamar Nomani; Julie Davila; Rehman Khan, Mastering Cloud Security Posture Management (CSPM): Secure multi-cloud infrastructure across AWS, Azure, and Google Cloud using proven techniques , Packt Publishing, 2024.

\bibitem{b6} A. Celesti, F. Tusa, M. Villari, and A. Puliafito, “How to enhance cloud architectures to enable cross-federation,” in Proc. 2010 IEEE 3rd Int. Conf. Cloud Comput., 2010, pp. 337–345. doi: 10.1109/CLOUD.2010.46.

\bibitem{punn2025privacy}
V.~Punniyamoorthy, A.~G.~Parthi, M.~Palanigounder, R.~K.~Kodali, B.~Kumar, and K.~Kannan,
``A Privacy-Preserving Cloud Architecture for Distributed Machine Learning at Scale,''
International Journal of Engineering Research and Technology (IJERT), 
vol.~14, no.~11, Nov.~2025.

\bibitem{b3} B. K. Malamuthu, V. S. Pandi, S. D, A. H. Jaber, J. Giri and R. Y. P, "Examining Multi-Cloud Architectures to Provide Enterprises with Resilient, Scalable, and Economical Cloud Computing Solutions," 2025 International Conference on Engineering Innovations and Technologies (ICoEIT), Bhopal, India, 2025, pp. 970-975, doi: 10.1109/ICoEIT63558.2025.11211532.



\bibitem{nagpal2024cicd}
A.~Nagpal, B.~Pothineni, A.~G.~Parthi, D.~Maruthavanan, A.~R.~Banarse, P.~K.~Veerapaneni,
S.~R.~Sankiti, and V.~Jayaram,
``Framework for automating compliance verification in CI/CD pipelines,''
International Journal of Computer Science and Information Technology Research (IJCSITR), 
vol.~5, no.~4, pp.~17--27, 2024. doi: 10.5281/zenodo.1425967.


\bibitem{b8}S. Gupta, M. Sundararamaiah, and G. Geeta, “Leveraging cloud-native data engineering for big data analytics,” in Proc. 2025 3rd Int. Conf. Adv. Comput. Comput. Technol. (InCACCT), 2025, pp. 976–979. doi: 10.1109/InCACCT65424.2025.11011292.


\bibitem{BigQuery}
J. Levandoski, G. Casto, M. Deng, R. Desai, P. Edara, T. Hottelier, A. Hormati, A. Johnson, J. Johnson, D. Kurzyniec, S. McVeety, P. Ramanathan, G. Saxena, V. Shanmugan, and Y. Volobuev,
``BigLake: BigQuery's evolution toward a multi-cloud lakehouse,''
in Proc. Companion Int. Conf. Management of Data (SIGMOD '24),
Santiago, Chile, 2024, pp.~334--346, doi:~10.1145/3626246.3653388.


\bibitem{Redshift}
A. Gupta, D. Agarwal, D. Tan, J. Kulesza, R. Pathak, S. Stefani, and V. Srinivasan,
``Amazon Redshift and the case for simpler data warehouses,''
in \emph{Proc. ACM SIGMOD Int. Conf. Management of Data (SIGMOD '15)},
Melbourne, Australia, 2015, pp.~1917--1923, doi:~10.1145/2723372.2742795.


\bibitem{Azure}
J. Aguilar-Saborit \emph{et al.},
``POLARIS: The distributed SQL engine in Azure Synapse,''
in Proc. Int. Conf. Very Large Data Bases (VLDB),
2020, pp.~3204--3216.


\end{thebibliography}
\end{document}